\documentstyle[aps,multicol,pre]{revtex}
\title{A correlation between the equilibrium and transport properties\\  
of intercalation systems.
}

\author{E.V. Vakarin  and J.P. Badiali
}
\address{Laboratoire de Electrochimie et Chimie Analytique\\
 ENSCP-UPMC, 11 rue P. et M. Curie, 75231 Cedex 05, Paris, France\\
}

\begin{document}
\maketitle
\begin{abstract}
Employing the lattice gas  model, combined with the 
linear elasticity theory, a correlation  
between the equilibrium and transport properties of intercalated 
species is investigated. It is shown that the major features of the 
intercalation isotherms and the concentration dependence of the chemical 
diffusion coefficient can be well understood in terms of the change of the 
host volume in the course of intercalation. Theoretical predictions are 
compared to the experimental observations on $PdH_x$, $Li_xWO_3$ and 
$Li$-graphite systems.  
\\   {\it PACS numbers:} 68.35.Ct, 68.35.Rh
\end{abstract}

\begin{multicols}{2}
\section{Introduction}
Intercalation processes find their application in many technologically 
important domains, such as the design of   
hydrogen-storage systems\cite{HS}, rechargeable high-energy  batteries,
electrochromic devices, (see Ref.\cite{rev} for a review), and 
superconductors \cite{sup}. 

The insertion process can be viewed (at least, in some aspects)
as an adsorption of guest particles
on the host lattice. For charged particles
the ionic charge inside the matrix is compensated
by the electrons. For that reason the electrochemical intercalation is also 
similar to a three-dimensional adsorption of
neutral species. Based on this analogy the intercalation is traditionally
described within the lattice gas (LG) model. In this approach all the
properties (intercalation isotherm or capacity-concentration dependence)
are connected with an ordering the guest on different adsorption
sites of a rigid host lattice - 
the  {\it configurational} transitions.   

However, the host response to the accommodation of the guest species
is not negligible. For instance, 
the insertion induces the stress
into the host matrix. This may lead to segregation effects \cite{seg} or
even to instabilities\cite{inst} of the host-guest system. Also a loading
path is shown\cite{loading} to influence the guest uptake efficiency.  
Due to this the host may undergo an expansion
or local distortion. Typical examples are the hydrogen sorption
by metals \cite{loading,bar,zol,zhang}, or the intercalation of $Li$ ions 
into layered materials \cite{ber,thomp,Dahn1,Dahn2}.
Quite often the host undergoes  structural transformations 
\cite{Mat,D1,D2,D3} due to intercalation.  
These are the {\it structural} transitions.
 In such cases the standard LG approach also
operates with several sublattices \cite{ber,D3} 
corresponding to each configuration of the host. 
However, such a restructuring suggests that the elastic effects should be 
taken into account. A general thermo-mechanical theory of the 
stress-composition interaction is developed by Larche and Cahn \cite{LC}. 
This implies the existence of a coupling between the elastic properties of 
the host material and the structure and/or the dynamics associated with the 
guest species. In other words the configurational and structural 
transitions should be considered to be coupled. 
In our previous papers we have investigated such a coupling for
two-dimensional \cite{PRL} as well as for three-dimensional
systems\cite{PRBd,EA,JPCB}. The main advantage of our approach is its 
capability in describing various (microscopically different) intercalation 
systems within a common framework. Despite of the fact that some microscopic 
details are omitted, the theory agrees well with experimental data 
on layered $Li_xTiS_2$ and crystalline $Li_xWO_3$ and $Na_xWO_3$ compounds
at equilibrium conditions. 

In this study we focus on a correlation between the equilibrium properties
(isotherms and differential capacity) and the concentration dependence
of the chemical diffusion coefficient.  Our purpose is to determine 
the impact of the thermodynamics to the kinetic properties.  
Following our previous works we combine 
the LG model, describing the configurational transitions and the linear 
elasticity theory that accounts for the loading mechanism and the volume 
dilatation in the course of the structural transitions (expansion or 
restructuring). This allows us to derive an effective chemical potential,
involving the concentration dependent stress and strain fields. Then,
assuming that the diffusion flux is proportional to the gradient of the
chemical potential, we investigate the behavior of the chemical diffusion 
coefficient.

\section{Model}
\subsection{Microscopic formulation}
The host material is described as a three-dimensional lattice of adsorbing
sites with their positions given by a set vectors $({\bf r}_i)$. Due to the
elastic properties of the real host
each site (for instance, an interstitial site) of this "auxiliary" lattice  
may deviate from its equilibrium position ${\bf r}_i^0$, such that we deal 
with the displacements ${\bf u}_i={\bf r}_i-{\bf r}_i^0$. Therefore, the 
host properties are described by the Hamiltonian $H_H[\{{\bf u}_i\}]$. Note 
however that a connection between the elastic properties of the real matrix 
and those of the adsorbing lattice, is not straightforward\cite{PRBd,EA}. 

A distribution of the intercalants on the host sites is given by
a set of occupation numbers $\{t_i\}$,
with $t_i=0$ or $t_i=1$. The guest subsystem is characterized by the chemical
potential $\mu$ and the nearest neighbor interaction parameter $W$ 
(at equilibrium positions).
For the electrochemical insertion,
 chemical potential $\mu$  gives a
deviation of the electrode potential $-eV$ from its standard value $E_0$.
In this way an arrangement of the intercalants on a rigid host
lattice is governed by the LG Hamiltonian
\begin{equation}
H_G=W\sum_{ij}t_it_j-\mu\sum_{i}t_i
\end{equation}
that describes the configurational transitions of the intercalated species.
These could be the droplet formation for attractive interactions ($W<0$)
or the order-disorder transition for repulsive interactions ($W>0$). In the 
latter case one introduces the sublattice concentrations as
appropriate for the symmetry of a given system.
 
The coupling between the host and the guest is given by Hamiltonian
$H_C[\{{\bf u}_i\},\{t_i\}]$
which takes into account a dependence of the binding energy on the site
displacement  and also the pairwise interaction between the
guest particles through the host lattice.
 The overall Hamiltonian is now written as
\begin{equation}
H=H_H[\{{\bf u}_i\}]+H_G[\{t_i\}]+H_C[\{{\bf u}_i\},\{t_i\}]
\end{equation}
The free energy $F$ corresponding to the above Hamiltonian is given by
\begin{equation}
F=F_H+F_G(x)+F_C(x)
\end{equation}
where $x$ is the intercalant concentration.
Here $F_H$ is the host free energy in the absence of intercalation, $F_G$ is
the guest free energy in a case of the rigid host lattice. The latter term
can be calculated using the mean field approximation to obtain
the well-known relation\cite{rev,ber} for the chemical potential
\begin{equation}
\label{mu0}
\mu_0(x)=-eV-E_0=cWx+\frac{1}{\beta}\ln \left(\frac{x}{1-x} \right)
\end{equation}  
where  $c$ is the coordination number
of the host lattice and $\beta=1/kT$.
The coupling term 
$\beta F_C=-\ln(\langle \langle e^{-\beta H_C}\rangle_{({\bf u}_i)} 
\rangle_{(t_i)})$ requires the averaging over the displacements and
the occupation numbers, calculated with the reference terms
$H_H$ and $H_G$. In fact, this is an infinite series including the 
correlations of all orders in the reference state. The main problem
is to specify $H_C[\{{\bf u}_i\},\{t_i\}]$ coherently
with the host symmetry and elastic properties. It is known that 
real host materials have rather complicated elastic properties (e.g. a 
strong anisotropy). Therefore, only some simplified  
model calculations are expected to give tractable results. On the other 
hand, such predictions (e.g. a rigid plane model\cite{plane})
do not exhibit quantitative agreement with experimental data. 
\subsection{Approximation}
The perturbative scheme developed previously \cite{PRBd,EA} has shown 
 that the coupling term is concerned with a concentration dependence of the 
 host response to the intercalation. This involves at least  
 two effects. First is a renormalization\cite{rev,Kalik} of the net pair 
 interaction between the intercalants due to their indirect interaction
through the matrix. The second is  a change of the host volume upon 
insertion of the guest species. Depending on the host nature, a  stress 
field may result if the lattice is not totally free to relax. In this 
situation it seems reasonable to estimate the coupling term based on the 
continuum theory of elasticity with the concentration dependent stress and 
strain fields. Then the host-guest free energy is a sum of the lattice 
gas and elastic part 
\begin{equation}
\label{fx}
F(x)=F_{LG}(x)+F_{el}(x)
\end{equation}
where $F_{LG}(x)$ is the configurational (lattice gas) part, in which the
pairwise interaction is renormalized\cite{rev} due to the interaction
through the matrix. This gives the chemical potential $\mu_0(x)$ 
(see eq.~(\ref{mu0})) with a new interaction constant $W$.
The elastic part is approximated by the free energy of a strained isotropic
body under a loading stress $\sigma(x)$. Since the strain is assumed 
to be purely dilatational, we operate with traces $\varepsilon$ and
$\sigma$ of the corresponding tensors. 
\begin{equation}
F_{el}(x)=\frac{\Lambda}{2}\varepsilon(x)^2-\sigma(x)\varepsilon(x)
\end{equation}
with $\Lambda$ being  the  effective elastic constant, independent of the
concentration. 
The total stress $S=S(x)$ is given by
\begin{equation}
S(x)=\frac{dF_{el}}{d\varepsilon(x)}=
\Lambda\varepsilon(x)-\sigma(x)
\end{equation}
Therefore, we have two stress contributions. The internal, or self-stress
$\Lambda\varepsilon(x)$ corresponds to the host reaction to the guest 
insertion. The second term $\sigma(x)$ describes a loading 
procedure, that may include the sample clamping or other effects
which are not directly related to the strain. It is important that,
in general,
$\sigma(x)$ is a function of $x$ (not a function of $\varepsilon(x)$).
For instance, if the sample is clamped such that
$\varepsilon(x)=0$, then we have a stress accumulation proportional to
the concentration $\sigma(x) \propto x$.  
\section{Equilbrium properties}
The guest chemical potential $\mu(x)$ is given by the concentration 
derivative of the total free energy.
\begin{equation}
\label{muel}
\mu(x)=\mu_0(x)+ S(x) \frac{d\varepsilon(x)}{dx}-
\frac{d\sigma(x)}{dx}\varepsilon(x)
\end{equation}
Here the second term involves the so-called chemical expansion
coefficient $d\varepsilon(x)/dx$, while the last term is associated
with the loading path. It is seen that the intercalation level depends on an 
interplay of the internal stress and the loading stress. The latter could be 
small, but its concentration  derivative is not necessary small, so that
the loading path may induce serious consequences\cite{loading}. In 
particular, for a given material ($\Lambda$) and a suitable loading path 
$\sigma(x)$, there may be a cancellation of the last two terms 
in eq.~(\ref{muel}) in a given domain of $x$. This explains why in some cases
the purely configurational description ($\mu(x)=\mu_0(x)$) works well. 

For comparison to other theoretical approaches it is convenient to express 
the strain function in terms of other relevant quantities. The strain can 
be measured as a volume dilatation or as a change of the interlayer spacing
in the course of insertion. If $V(x)$ is a composition dependent sample 
volume (or interlayer spacing for layered compounds), then by definition 
\begin{equation}
\varepsilon(x)=\frac{V(x)-V(0)}{V(0)}=\delta p(x)
\end{equation}
where $\delta= [V(1)-V(0)]/V(0)$ is the  relative volume variation
and $p(x)=[V(x)-V(0)]/[V(1)-V(0)]$ is a modulating function. The latter
varies in between $0$ and $1$. If $p(x)$ is a linear function, then the 
system is said to obey the Vegard's law\cite{Vegard}. 

If the host-guest system 
forms a solid solution, then the guest partial molar volume $V_m(x)$ is 
related to the total volume 
\begin{equation}
V(x)=xV_m(x) +V(0), 
\end{equation}
where $V(0)$ is the initial host volume (at 
$x=0$). Then we can find the following relation for the strain,  
the volume variation and partial molar volume 
\begin{equation}
\varepsilon(x)=\frac{V(x)-V(0)}{V(0)}=x\frac{V_m(x)}{V(0)}
\end{equation}
If the loading is composition independent $\sigma(x)=\sigma$ and the
sample volume changes linearly ($V_m=const, \varepsilon(x)=xV_m/V(0)$,
like in the $PdH_x$ $\alpha$-phase), 
then we recover the well-known result\cite{zhang}
\begin{equation} 
\mu(x)=\mu_0(x) -\left(\sigma -\Lambda x \frac{V_m}{V(0)}\right)
\frac{V_m}{V(0)} 
\end{equation}
The term liner in the concentration can be viewed as an additional
interaction which can be combined with $cWx$ in $\mu_0(x)$. Although
this additional term is proportional to the elastic constant $\Lambda$,
it is physically different from the interaction of elastic 
dipoles\cite{rev,Kalik} which is already absorbed into $W$ (see the 
discussion after equation (\ref{fx})). This interaction takes place even if 
the sample is perfectly clamped ($\varepsilon(x)=0$), while our term 
represents a cooperative effect due to  $\varepsilon(x) \ne 0$.
 
In order to analyze the role of elastic effects in the phase behavior of the 
guest species it is instructive to consider a simple example of vanishing 
loading stress $\sigma(x)=0$, $S(x)=\Lambda \varepsilon(x)$. In this case 
there is no difference between the volume expansion or contraction. 
Then we arrive at
\begin{equation}
\mu(x)=\mu_0(x)+\bar\Lambda p(x)\frac{dp(x)}{dx}
\end{equation} 
where $\bar\Lambda=\Lambda \delta^2$. In addition, $p(x)$ is taken to be
that in the so-called layer rigidity model\cite{Vegard} which describes a
deviation from the Vegard's law in layered compounds
\begin{equation}
p(x)=1-(1-x)^q
\end{equation}
where $q$ is related to the rigidity of the host layers. 
The first order phase 
transition (of liquid-gas type) is manifested by a singularity in the
differential capacity $C=dx/d(\beta \mu)$
\begin{equation}
\label{C}
\beta C=\left[
\frac{d\mu_0(x)}{dx}+\bar\Lambda \left(\frac{dp(x)}{dx}\right)^2+
\bar\Lambda \frac{d^2p(x)}{dx^2}
\right]^{-1}
\end{equation} 
It is clear that for $q<1$ both derivatives of $p(x)$ are positive and the 
only possibility for the critical behavior is the case of attractive 
interactions $W<0$, i.e. when the configurational part  
\begin{equation}
\beta \frac{d\mu_0(x)}{dx}=\frac{1+\beta c Wx(1-x)}{x(1-x)}
\end{equation} 
becomes negative. Then the last two terms in the denominator of (\ref{C})
rescale the critical temperature and shift the critical composition from
$x=1/2$.  For $q>1$  the term $d^2p/dx^2$ becomes negative and  
the criticality may appear due to the elastic effects even if $W=0$. The
coexistence curves for this case, obtained by a numerical analysis of the 
divergence in $C$, are plotted in Fig.~1. It is seen that the critical 
temperature (the maximum) increases with increasing $q$. Simultaneously the 
critical concentration decreases. The elastic effects break the 
hole-particle symmetry of the problem and consequently the curves are not 
symmetric and are not centered around $x=1/2$.

In our previous works 
\cite{PRBd,EA,JPCB} we have analyzed another extreme
case of vanishing total stress $S(x)\to 0$ and the loading stress 
proportional to the concentration $\sigma(x)=\sigma_0x$. Then
\begin{equation}
\label{msf}
\mu(x)=\mu_0(x) -\gamma p(x)
\end{equation} 
where $\gamma=\delta\sigma_0$. Note, however 
that, for $\Lambda=const$,  the above result is valid only for a linear 
behavior of $p(x)=px$, while  is an approximation for an arbitrary
$p(x)$. In reality the elastic constants could depend on the concentration,
for instance the bulk modulus of $Pd$ is reduced\cite{zol} up to $20 \%$ 
due to the hydrogen sorption. Thus one can easily imagine a situation when
$\Lambda(x)\varepsilon(x)\approx \sigma(x)$ and then the approximation
(\ref{msf}) is applicable for a nonlinear $p(x)$. In general, for a 
non-Vegard's behavior and the linear loading $\sigma(x)=\sigma_0x$ we deal 
with 
\begin{equation}
\label{muf}
\mu(x)=\mu_0(x)+
\left[
\bar\Lambda p(x)-\gamma x 
\right]
\frac{dp(x)}{dx} -\gamma p(x)
\end{equation} 

The strain can be measured as a volume dilatation or
as a change of the interlayer spacing during the transitions between 
different phases (staging in graphite, restructuring in $Li_xWO_3$, 
$\alpha -\beta$ transition in $PdH_x$, etc). In any case the lattice
parameters do not obey the linear Vegard's law \cite{Vegard}, exhibiting a 
sharp change near the transition compositions $x_n^0$. This can be described 
by the following approximation\cite{PRBd,EA,JPCB}.    
\begin{equation}
\label{p}
p(x)=\frac{1}{2}\left[1 + \sum_n p_n\tanh[\alpha_n(x-x_n^0)]\right]
\label{p}
\end{equation}
where $n$ counts the number of phase boundaries. In $Li_xWO_3$,
$n=1,2$  
corresponds to monoclinic-tetragonal and tetragonal-cubic
transitions, respectively. For $Li$-graphite systems, $n$ corresponds
to the stage transfer boundaries\cite{Dahn1,Dahn2}.  
The set of rigidity 
parameters $\alpha_n$ controls
a local slope and 
$p_n$ are the weights corresponding
to each phase, such that $\sum_n p_n=1$. This is consistent with 
experimental observations\cite{Mat} indicating that the 
structures are not completely pure, but contain some features that indicate 
a mixing of phases. The present form of $p(x)$ corresponds to continuous
structural transitions, but can be easily modified to take into account
the jump-like behavior (like for the staging \cite{rev}). Nevertheless, in 
that case the concentration derivatives of $p(x)$ would be singular, 
inducing the singularities in the thermodynamic quantities (as it should be
in the neighborhood of a phase transition). Note however
that the experimental dependencies are usually smoothed due to a finite
concentration resolution. Then it is difficult to distinguish between the 
step-wise variation and a sharp (but continuous) transition. For that reason
the criticality criteria, determined on the ground of rigorous statistical
mechanical arguments ($\mu(x)$ loops or a divergent capacity 
$C(x)$), only approximately conform to experimental data. 
Therefore we stay with the continuous $p(x)$, performing the fitting to
the experiment.

Following this methodology, 
in Fig.~3 we display  the intercalation isotherm (inset) and
the capacity curve, comparing them to the experimental data\cite{Mat}
for $Li_xWO_3$. The fitting (eq.~(\ref{muf})) is performed assuming
that the effective pair interaction inside the matrix is repulsive
for any $x$. The curve modulation results from the strain behavior
$p(x)$, associated with the volume dilatation.   
As is discussed previously\cite{EA,JPCB}, the peculiarities 
(inflection points and the peaks in $C(x)=dx/d(\beta \mu)$) mark the 
boundary between different host symmetries. From the experimental point of 
view, the peak height is usually associated with a transition sharpness.
Nevertheless, as we have discussed above,  
the critical behavior of the model itself should be studied separately.
\section{Chemical diffusion coefficient}
In many cases there are two mobile species inside host matrices:
the host electrons and the guest particles. Therefore, the transport
is characterized by their flux densities ${\bf J}_e$ and ${\bf J}_G$,
related to the gradients of the corresponding chemical (or electro-chemical)
potentials $\mu_e$ and $\mu_G$.  
\begin{equation}
{\bf J}_e=L_{ee}\nabla \mu_e+L_{eG}\nabla \mu_G
\end{equation}
\begin{equation}
{\bf J}_G=L_{GG}\nabla \mu_G+L_{Ge}\nabla \mu_e
\end{equation}
where $L_{ab}$ are phenomenological transport coefficients, which are
scalar quantities if the host is spatially isotropic. In matrices with 
a metallic conductivity the electrons are much more mobile than the guest 
species. Due to this the guest ionic charge
is compensated by the electrons. For the same reason 
the electrons reach a uniform (equilibrium) distribution 
much faster than the intercalants ( $\nabla \mu_e=0$
on the time scale  when $\nabla \mu_G \ne 0$).  In addition, usually
$L_{GG}>> L_{eG}$ \cite{rev}. Therefore one may focus on the guest flux
\begin{equation}
\label{flux}
{\bf J}=-M(x)x\nabla \mu=-D(x)\nabla x
\end{equation}
where $M(x)$ is the mobility coefficient that must contain a blocking
factor $M(x)=D_0(1-x)$, determining the concentration-dependent chemical
(or collective) diffusion coefficient $D(x)$. A comprehensive review of
theoretical approaches to the determination of the chemical diffusion 
coefficient can be found in Ref.~\cite{Gomer}. In the simplest approximation
we deal with 
\begin{equation}
\label{Ddef}
D(x)=D_0x(1-x)\frac{d \beta \mu}{d x}=D_0x(1-x)/C(x)
\end{equation}
Under the assumptions above, the intercalant transport is described
by the diffusion equation with an effective diffusion coefficient
$D(x)$. The latter requires an information on the intercalation isotherm
$\mu(x)$. Then, based on the equilibrium properties, at a given concentration
gradient, one, at least in principle, can solve the kinetic problem.

Note, that such a simple scheme should be modified when the host 
conductivity changes remarkably upon the intercalation.  Then the 
electronic impact should be taken into account. Also, we do not consider
other driving forces, like external field or stress gradients, assuming
that the guest concentration, $x$, is the only independent variable. And 
finally, near the phase coexistence (e.g, staging) a system becomes 
non-uniform  because of the phase boundaries. Then  
the formulation of the 
kinetic problem must be coherent with the theory of critical phenomena
(see \cite{diff} for a recent review).

Starting from (\ref{muel}) and (\ref{Ddef}) we obtain
\begin{equation}
D(x)=D_0(x)+D_{el}(x)
\end{equation}
where the first term is the standard LG part
$D_0(x)=D_0x(1-x)\frac{\partial \mu_0}{\partial x}$ and the elastic 
contribution is given by 
\begin{eqnarray}
\label{Dres}
D_{el}=B(x)
\left[
S\frac{d^2\varepsilon}{dx^2}
+\Lambda \left(\frac{d\varepsilon}{dx}\right)^2 
-2\frac{d\varepsilon}{dx}\frac{d\sigma}{dx} 
-\frac{d^2\sigma}{dx^2} \varepsilon
\right]
\end{eqnarray}
with $B(x)=D_0x(1-x)$.
The diffusion coefficient involves several
competing factors (strains, stresses and their  
concentration derivatives). In general it is not trivial to see whether $D$ 
increases or decreases with $x$. The situation is even more complicated for 
systems in which the lattice spacing  does not obey the linear Vegard's 
law, but has  inflection points separating different phases (e.g. staging 
in graphite or restructuring in $Li_xWO_3$). Then the derivatives above may 
change sign with the concentration. Also we see again that the loading path 
has a significant contribution.

For $PdH_x$ ($\alpha$-phase) we recover the well known result
\cite{zol,zhang}. 
The concentration induced 
internal stress $S_0(x)=\Lambda x V_m/V_0$ increases 
the diffusion  
\begin{equation}
D= D_0(x)+B(x)\Lambda  \left( \frac{V_m}{V_0}\right)^2,
\end{equation}
where $D_0(x)$ is the stress free contribution (it 
corresponds to $\mu_0(x)$, i.e. the lattice gas description, see above). 
Note that the non-local stress effects\cite{bar,zol} are not discussed here.

Since, according to (\ref{Ddef}), the diffusion coefficient is just 
an inverse of $C(x)$, $D$ as a function of $x$ has minima (Fig.~3), 
corresponding to the peaks in the capacity curve. 
For $WO_3$ (Fig.~3) the theory works well near the first minimum,
but overestimates the diffusion at higher $x$. 
Assuming that eq.(\ref{Ddef}) is valid, we have also inverted the capacity 
data (Fig.~2)  in order to demonstrate that  there is no "hidden" errors 
in the fit for $C(x)$. Since the equilibrium characteristics 
(Fig.~2) are predicted with a reasonable accuracy, then one of the source 
for the discrepancy could be our simplified assumptions on the mobility.    
For instance,
one might suppose that eq.(\ref{Ddef}) does not work for higher $x$ 
because the hoping $D_0$ depends on the elastic properties.  By the analogy 
with the two-dimensional diffusion\cite{Gomer}, $D_0$ is related to the 
lattice spacing. The latter decreases with $x$ for $WO_3$\cite{D1}, so that 
the resulting $D(x)$ would also decrease. Nevertheless the volume change
does not exceed several percents\cite{rev,D1}. This cannot explain the two 
orders of magnitude difference in Fig.~3.   
The lattice  anisotropy  seems also to be irrelevant because the final (high 
$x$) structure is cubic, so the isotropic approximation for the diffusion 
and elastic properties is reasonable. 

It is known\cite{cond} that
$Li_xWO_3$ changes its electronic conductivity near the structural
instability composition. Then the ionic transport must correlate with the
transport of the neutralizing electrons. 
Similar effects occur in 
superconductors\cite{sup}. Therefore, the electronic mobility impact
(that is absent from our approach)
to the current should be taken into account by including the additional
driving force $\nabla \mu_e=d\mu_e/dx \nabla x$. This contribution can be
included into our approach, assuming some model for the electronic 
structure, e.g. - the rigid band model. This would correct our predictions 
for small $x$ (according to \cite{cond}, for $x \le 0.1$), but our high-$x$ 
estimation would remain unchanged. We believe that the electronic effects
are not responsible for the decrease of $D(x)$ at high concentrations.
Based on the fact\cite{cond} that the electrical properties of $Li_xWO_3$ 
become more and more metallic with increasing $Li$ content, we may expect
that our approximation $\nabla \mu_e=0$ becomes more reliable with increasing
$x$. Moreover, if the conductivity effects were important, then the ionic 
diffusion would increase like this occurs in amorphous $WO_3$ 
films\cite{amorphous}.

On our opinion we deal with an interplay of several facts. Although
the model works well in predicting the equilibrium properties, nevertheless
it is probably too simple for the description of the kinetics. In fact,
the transport is assumed to take place in an infinite lattice through the 
hoping of neutral species ($Li^++e$), ignoring the kinetics other relevant 
phenomena, like the charge transfer (metal/host, guest/host), the exclusion 
(permselectivity) effect\cite{PRBd}, the formation of the passive layer at
the host/electrolyte boundary, etc. On the other hand, the experimental data
on the diffusion are extracted from direct measurements employing a 
theoretical model, which could differ (in some details) from the one
introduced here. Concerning the order of magnitude estimations, it should be 
noted that the experimental data for the same substance are usually differ
significantly depending on the sample preparation, its size and the scanning 
time interval. Therefore, the quantitative description of the ionic 
transport is a delicate problem requiring mutual theoretical and 
experimental efforts.  
Nevertheless, our estimation of the elastic effects is qualitatively
correct in predicting that the system characteristics 
changes coherently with the strain $p(x)$, and therefore,
positions of the minima (Fig.~3) $x_n^0$ are predicted correctly, implying a 
correlation between the $D(x)$ behavior and the strain development. 
 
Similar situation takes place for $Li$-graphite. 
In Fig.~4 the normalized interlayer spacing for  graphite in the
presence of intercalated $Li$  is plotted as a function of the guest 
composition. The fitting is preformed using eq.~(\ref{p}) under a suitable
choice of the parameters $p_n$, $\alpha_n$, $x_n^0$. 
The experimental data are taken from Refs.~\cite{Dahn1,Dahn2}.
Our fit is 
rather reasonable, "catching" the fact that the stage transfer compositions
$x_n^0$ approximately correspond to the inflection points in $p(x)$,
where the average interlayer spacing changes sharply. Note that we do not
discuss the fine structure of the phase diagram\cite{Dahn1}, such as
dilute and liquid-like phases for the same stage. These features are 
related to the the guest in-plane ordering which is indistinguishable in
the behavior of the interlayer spacing.

Having a reliable
approximation for $p(x)$ we calculate the diffusion coefficient in the 
framework outlined above. The fitting to the experimental data \cite{levid} 
is shown in Fig.~5. It is seen that the diffusion slows down significantly
near the stage transfer compositions $x_n^0$. Although the magnitude of 
$D(x)$ is close to the experimental results, the width of the minima is
underestimated. Note that the fitting is not optimized, that is, we did not 
try to find an optimal set of the parameters, which gives equally good 
agreement for the isotherm $\mu(x)$ and $D(x)$. 
Nevertheless, as in the 
case of $Li_xWO_3$ (Fig.~3), there is a clear correlation between the 
diffusivity and the strain behavior.  
 
\section{Conclusion}  
In summary, our approach implies that the main equilibrium and transport
features of the intercalation systems 
(which differ in their microscopic details) can be well understood in terms 
of a concentration dependence of the hydrostatic parts
of the stress and strain fields, associated with the internal and loading 
effects\cite{loading}. 
It is shown that the elastic effects may induce the critical behavior even
if the "direct" interaction between the guest species is absent.   
Although the ionic transport is supposed to be equivalent to the diffusion
of neutral (ion plus electron) species, the approach is flexible enough 
to incorporate other transport mechanism (migration, or electronic mobility 
effects).  

The theory gives a quantitative description of
different insertion processes, involving the volume expansion 
($Li_xTiS_2$ (see\cite{PRBd}) , $\alpha-PdH_x$), staging ($Li$-graphite) or 
restructuring ($Li_xWO_3$). For all these processes the theory implies a 
correlation between the intercalation isotherm and a concentration 
dependence of the diffusion coefficient. The latter exhibit a set
of characteristic minima, related to the boundaries between different
phases (like  different symmetry phases of $Li_xWO_3$, $\alpha-\beta$
transition in $PdH_x$, etc). However
the experimental dependencies are usually smoothed due to a finite
concentration resolution.  For that reason
the criticality criteria, determined on the ground of rigorous statistical
mechanical arguments (e.g. $D(x)=0$ at the transition concentrations) only 
approximately conform to the experimental data, exhibiting a sharp (but 
finite) decrease of $D$.   

For a non-Vegard's strain variation the diffusion coefficient is
a nonlinear (and nonmonotonic) function of the concentration
(\ref{Dres}). Therefore the diffusion equation for the concentration 
profile would be strongly nonlinear. Then one can expect a rather
complicated space-time variation, including , for 
instance, oscillations\cite{loading} and other nonlinear effects.

Our results may have implication in various domains related to the
insertion process, like 
hydrogen sorption\cite{HS}, electrochemical intercalation, impurities in 
alloys, layered superconductors\cite{sup}, volume transitions in hydrated 
gels\cite{gel}, etc.

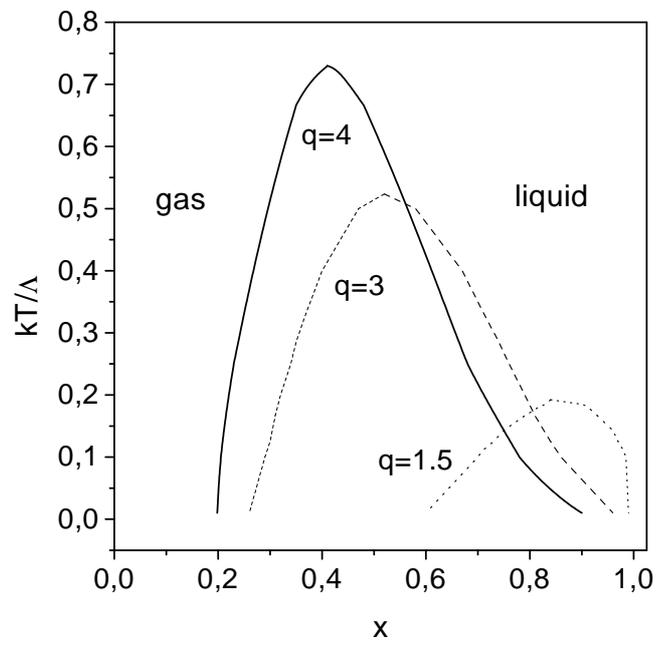
\begin{figure}
\caption{ Liquid-gas coexistence curves for the layer rigidity model
in the absence of the direct interaction $W=0$. 
}
\end{figure}

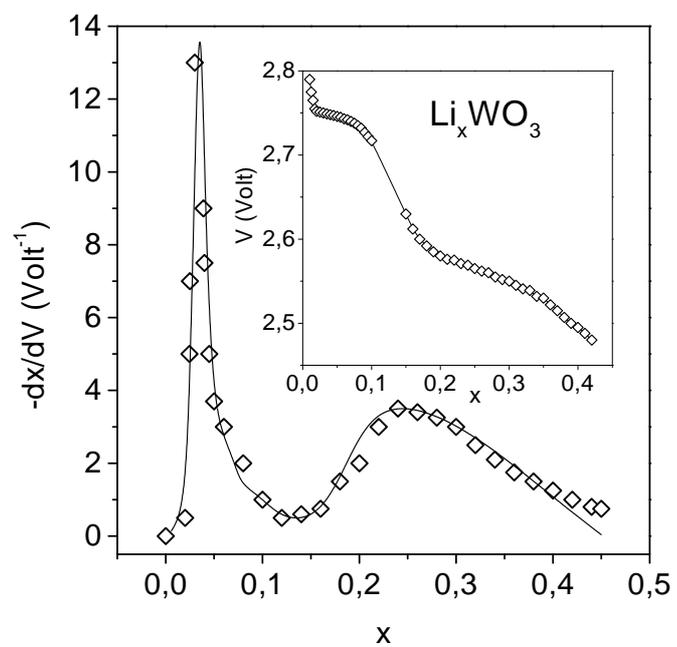
\begin{figure}
\caption{  Differential capacity and the voltage (inset) for
crystalline $Li_xWO_3$. The symbols correspond to the experimental
data [12]. The parameters $p_1=0.7$, $p_2=0.3$, $\alpha_1=\alpha_2=15$, 
$x_1^0=0.05$, $x_2^0=0.27$, $\beta \gamma=1.8$, $\beta \bar \Lambda=0.01$, 
$\beta W=2.2$. 
}
\end{figure}

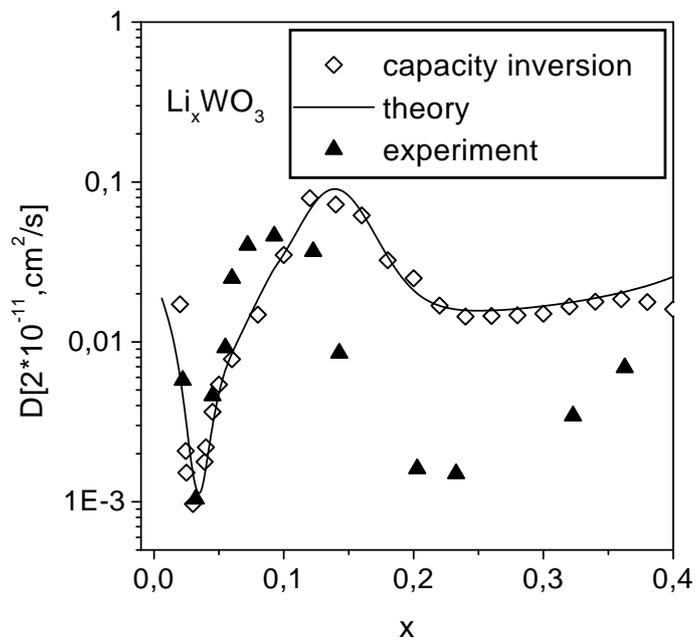
\begin{figure}
\caption{The chemical diffusion coefficient for
crystalline $Li_xWO_3$. The symbols (up triangles) correspond to the 
experimental data [24]. The parameters are the same as for the 
previous figure. 
} 
\end{figure}

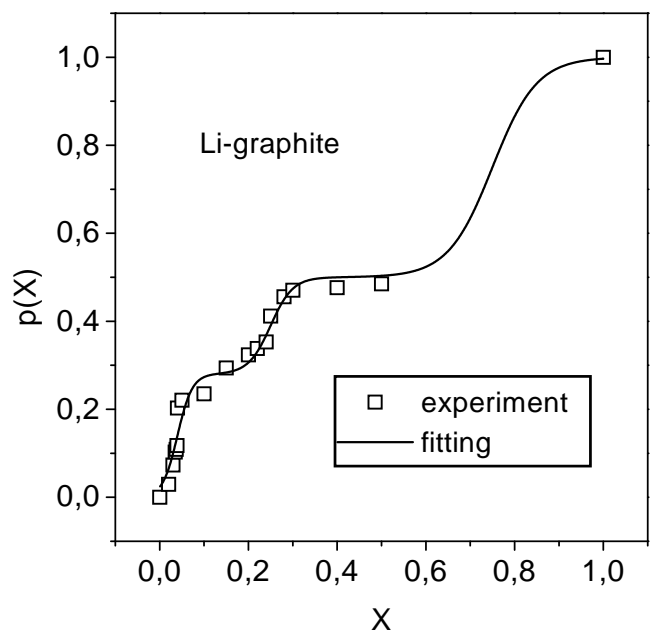
\begin{figure}
\caption{ The average interlayer spacing for $Li$-graphite. The experimental
data are extracted from [12,13]. The theoretical fitting is performed using
equation (19), where $p_1=0.28$, $\alpha_1=30$, $x_1^0=0.04$,
$p_2=0.22$, $\alpha_2=20$, $x_2^0=0.25$,
$p_3=0.5$, $\alpha_3=10$, $x_3^0=0.75$
 } 
\end{figure}

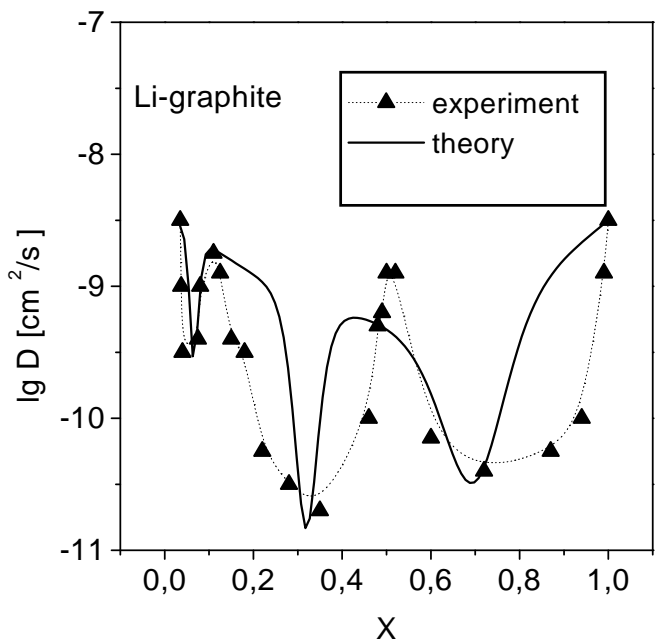
\begin{figure}
\caption{The chemical diffusion coefficient for $Li$-graphite. The 
experimental data are extracted from [31]. The fitting is done using 
equations (18) and (23) where $\beta c W=-0.1$, $\gamma=0.5$ $\bar 
\Lambda=0.1$   } 
\end{figure}
\end{multicols}
\end{document}